# Cooling down and waking up: Feedback cooling switches an unconscious neural computer into a conscious quantum computer

Andrew Bell
Eccles Institute of Neuroscience
John Curtin School of Medical Research
Australian National University
Canberra, ACT 2601
Australia

*Abstract:* This paper sets out a theory of how feedback cooling in the brain switches on consciousness. It explains how cooling reduces thermal noise to the point where macroscale quantum phenomena – crucially Bose–Einstein condensation and long-range coherence – can operate at body temperature. It takes the core idea from Stapp that mind and brain interact via some sort of oscillator and then focuses on a likely candidate: neuronal arrays identified by Stapp as cortical minicolumns. Feedback cooling allows amplifiers to act as refrigerators, and when applied to minicolumns it is suggested that the units perform like quantum accelerators, solid-state devices devised to supercharge standard computers. When the accelerator is idle, as in sleep, we have a neural computer operating unconsciously, but when feedback cooling is activated by thalamocortical loops, it produces a Bose–Einstein condensate, quantum computation, and consciousness. The model explains how macroscale quantum phenomena can operate in a warm and noisy brain, how and why consciousness evolved, and gives insight into puzzling unconscious states like sleepwalking. The model is testable, predicting that cold states in the brain are detectable by magnetic resonance thermometry.

*Abstract 184 words. Text 15,250 words + 126 references + 3 figures*

# 1. Introduction

Quantum mechanics and consciousness lie at the very frontiers of science and, given their intriguing parallels, there have been numerous attempts to yoke them together (Al-Khalili & McFadden, 2014; Anastopoulos, 2021; Atmanspacher, 2020; Chalmers & McQueen, 2023; Gao, 2022; Hameroff, 2022; Hameroff & Penrose, 2014; Josephson, 2019; Penrose, 2022; Schwartz et al., 2005; Stapp, 2007). Here a fresh approach is taken, building on the quantum mechanical ideas of Stapp. The work starts with his basic neural oscillator model (Stapp, 2017), and, bringing in related ideas from Eccles and others, creates a neurally specific model that is amenable to scientific test. All the elements in the new model have foundations in the literature, and the synthesis carries a number of wide-ranging implications for brain science. Among them is providing a specific handle for understanding how and why consciousness evolved.

At the core of the new model is a recognition of the power of feedback cooling to reduce noise in a system and allow macroscale quantum mechanical states to exist within a warm, wet, and noisy environment. The virtues of feedback cooling have been made use of in physical systems such as gravity wave detectors, but, despite one notable exception, it has not been considered in a biological context. The present paper explores that exception, and examines the possibility that feedback cooling in the brain underlies the generation of consciousness. The suggestion is made that cooling in the cortex leads to the creation of a unique macroscopic quantum object, our conscious mind.

Moreover, from an evolutionary perspective, it becomes easier to understand how a conscious state could emerge from an unconscious one. The unconscious neural network is, in terms of evolution, already in place, so all that is now needed is for additional cooling circuits to be added, which will have the effect of turning an unconscious animal – a neural computer – into an awake being with superior conscious powers. As will be shown, this is the role of the neural loops that connect thalamus and cortex, and which we know play a crucial role in waking us up in the morning. An animal that is able to develop effective feedback cooling circuits in its cortex has the potential to access quantum processing, and here it is described as the switching on of a powerful quantum computer. There is a seamless transition from neural computer to quantum computer because they both share one and the same piece of hardware: the cortical minicolumn. In one case the system operates at normal body temperature; in the second at much reduced effective temperatures.

All of this will be explained step by step, but in doing so it also helps to recognise that the transition between the two sorts of computers is in many ways similar to what novel devices called quantum accelerators aim to achieve. These commercially available quantum mechanical devices are designed as add-ons to conventional computers and use quantum mechanical processes to speed up computation. You can go online today and purchase such a device off the shelf (https://quantumbrilliance.com/). Remarkably, such devices do not require cryogenic temperatures in order to operate, although many of them do. The quantum accelerator from Quantum Brilliance is based on 5 quantum mechanical bits (qubits), and involves the entanglement of the nuclear spins from 5 nitrogen nuclei embedded within a diamond substrate. The device operates at room temperature and is available as a 19-inch rack-mountable unit (Doherty, 2021). Its ability to speed up standard computations is a prime motivator for this paper. Its mode of operation is perhaps analogous to how evolution has worked to ramp up the processing power of animal brains. Additionally, the speculation is made that, as an added bonus, quantum computation in the brain happens to be accompanied by a remarkable and unique state – consciousness.

Given the acknowledged advantage of quantum computers in terms of processing power, the central question is: might the conscious human mind really be an instance of a quantum computer? Not only is there extra processing power, but quantum mechanics also provides the unity of experience that has long attracted physicists such as Stapp, Penrose, and Bohm, neuroscientists like Eccles, theoretical biologists like Igamberdiev, and an increasing number of philosophers (see the reference list, in particular Gao (2022) and Adams and Petruccione (2020) for surveys). The macroscopic quantum mechanical entity is there all at once, just like our consciousness is (Simon, 2019).[1] Our conscious mind is able to take in a multitude of sights, sounds, sensations, thoughts, and memories, all available for instant access (Nagel, 2012).

---

[1] By consciousness this paper means phenomenal consciousness, laden with qualia, not access consciousness (Block 1995). A nice definition of consciousness is the *here-I-am-ness* of Dennett (2013). It is plainer English than the "something that it is like to be" of Nagel (1974). Further, Nagel's wording has difficulty capturing meditational states which are "not like anything" (Shear 2007, p. 700), whereas Dennett's formulation captures some of the universal I-ness which underlies the experience. Stapp (1993) defines it more wordily as "[t]hat luminescent presence of coming-into-beingness that constitutes our inner world of experience" (p. 234).

Quantum mechanics offers a solution to the longstanding and perplexing problem of how minds and bodies interact. But if the mind is quantum mechanical, a crucial question is how can it sustain itself against the decoherence of thermal noise? The plan of this paper is to supply a possible answer.

This paper elaborates on ideas first sketched in earlier work (Bell et al., 2022) and are here placed on a firmer footing.

## 2. Stapp and neural oscillators

The approach here starts from a suggestion by Stapp (2017) of how conscious experience could arise in a physical brain. Stapp proposes (his Appendix F) that there are quantum states in the brain which, akin to an oscillator, can be described by simple harmonic motion. If so, then it is possible for the mind to repeatedly interact with that oscillator. If the mind effectively observes the oscillator every cycle, it means that the quantum Zeno effect can come into play. Although the mathematical details can be found elsewhere (Misra & Sudarshan, 1977), essentially the Zeno effect means that cyclical timing is able to control the evolution of an underlying quantum system. Very rapid measurements of a system effectively slow down its evolution, freezing it in place; similarly, but in reverse, rapid interactions can also accelerate the evolution of a system towards another desired state – the inverse quantum Zeno effect (Altenmuller & Schenzle, 1993). A readable account of these effects, and how they relate to Stapp's view of consciousness, can be found in Laskey (2018). A related treatment of the quantum behaviour of a neural oscillator has been set out by de Barros (2012).

Stapp considers that conscious intentions, perhaps a mental picture, engage the inverse Zeno effect so as to produce the outcome we want in the world. But, as Laskey well asks (p.152), where specifically in the brain might these quantum effects occur? Stapp suggests that the locus resides in "cortical minicolumns", assemblies of roughly 100 neurons filling a volume some 80 μm across and 1500 μm deep (Buxhoeveden & Casanova, 2002).

Based on its diameter and other parameters in the literature, Stapp calculates (p.125) that for a quantum oscillator rotating at 20 Hz (a mid EEG frequency), then, to sustain its quantum state, probing actions would need to occur about once per millisecond (some 1000 Hz), which he regards as a reasonable neurophysiological figure.

This simple model provides a good starting point, although its plausibility has been questioned. In particular, to keep the system coherent, feedback from the brain's electrical or magnetic fields onto ion channels at synapses is necessary, and the details of such a process are missing. The biggest factor here is environmental decoherence, which arises because of thermal noise in the brain. While 1 ms might be an appropriate timescale neurophysiologically, it is immensely longer than the timescale over which quantum interactions normally occur. Thus, Tegmark (2000b) calculates that at room temperature the decoherence time is about $10^{-20}$ s and on this basis comes to the conclusion that "the brain is probably not a quantum computer".[2] Stapp believes his theory can overcome this objection (Stapp, 2015), but a question mark remains. In discussing the matter, Laskey (2018) has said that a more neurologically plausible model would be helpful, and this paper aims to supply one.

The approach taken here focuses on cortical minicolumns, looking in more detail at their internal structure and possible behaviour. In particular, we note the conclusion of Buxhoeveden and Casanova (2002) that "the minicolumn seems to be the most basic and consistent template by which the neocortex organises its neurones, pathways and intrinsic circuits" (p.946) and in this context we suggest that one of the basic functions is to generate internal feedback between efferent and afferent nerve pathways. With a positive feedback circuit in place, it is natural to expect that, with gain above unity, continuous oscillation would occur, and this, we suggest, is the core of Stapp's fundamental oscillator. In other words, the augmented model sees electric currents circulating around a miniature coil of dendrites in the minicolumn, and this sets the scene for novel quantum mechanical effects. The first thing to recognise is that such a coil will generate a tiny magnetic field and this has implications for long-range quantum states.[3] The second aspect, and the one emphasised

---

[2] A stronger claim is made in Tegmark (2014), pp. 205–208, in a section headed "Why your brain isn't a quantum computer".
[3] It also has implications for how the brain, in particular of birds, may be able to detect tiny magnetic fields. The neural mechanisms behind long-range bird navigation is largely unknown, and it remains a puzzle as

here, is that the circuit could also form the basis of feedback cooling, in which, following known physical principles, the effective temperature of a system can be reduced to a level where macroscale quantum mechanical phenomena appear. That is, fermions pair up to become bosons, as happens with superconductivity, from which point long-range order, perhaps over the whole of the cortex, may appear. This sudden transition, it is suggested, forms the basis of conscious experience. When we wake of a morning, feedback cooling switches on, and a coherent conscious state rises above the noise of unconscious neural activity.

Before going further, it is worth noting that a similar idea – again focusing on cortical minicolumns as the basis units of consciousness – was made some years earlier by the neurophysiologist Eccles and the neuroanatomist Szentágothai (Eccles, 1990; Szentágothai, 1979). Eccles called the dendritic cluster a "dendron", and when energised into a quantum mechanical unit, a "psychon", effectively an atom of consciousness (Bell et al., 2022). However, the psychon model did not take into account the effects of either feedback or temperature. Interestingly, both Eccles and Stapp espoused a dualistic picture of the world, but neither appears to have made reference to each other's work. In what follows, we build on this same anatomical unit and use the terms interchangeably.

## 3. Eccles and psychons

Eccles was perhaps the first to connect the classical and quantum realms in an explicit neurophysiological way. Inspired by the quantum ideas of Margenau (1984), Eccles pointed to a precise locus where the two realms meet: at assemblies of pyramidal cells in the neocortex where efferent and afferent nerve fibres come together (Eccles, 1990), the same anatomical unit as later identified by Stapp as a minicolumn. Eccles considered that mind is generated by some quantum process in the assemblies, calling each of the mental–physical units "dendrons", or when active, "psychons" (Eccles, 1990). Eccles proposed that

---

to how such small fields can be detected above thermal noise (Hong, 1995). But if a minicolumn can undergo cooling and it can be configured to detect surrounding fields, the problem seems more tractable. Heyers et al. (2007) report functional connections between the retina and the thalamus, which is consistent with possible cooling circuits directed towards a retinal detector. A not unrelated speculation is that minicolumns might be major sources of the magnetoencephalogram (since action potentials cannot contribute magnetically).

the units displayed quantum mechanical properties, work which was later elaborated in collaboration with physicist Friedrich Beck (Beck, 2001; Beck & Eccles, 1992). Eccles and Beck supposed that the quantum mechanical processes occurring within synapses caused the dendron to "light up" and become a psychon.

For his whole life, Eccles had been looking for how the mind could reside within neural tissue (Bell, 2022; Popper & Eccles, 1977). How could awareness, full of life and sound and colour, arise from a conglomeration of neurons? The answer he thought could come from combining the neuroanatomical findings of his colleague Szentágothai and the quantum mechanical ideas of Margenau and Beck.

Figure 1 illustrates, from Eccles (1990), the anatomy of the pyramidal cells within the cortical minicolumn or dendron. Crucially, there is a tangle of synapses within the structure and a rich connection of efferent and afferent pathways. Afferents are ascending nerve pathways which convey action potentials from peripheral sense organs to the brain, while efferents are descending pathways which deliver command signals from higher brain levels down to muscles so as to perform motor actions. There is more to the story, of course, but the important point is that any connection between afferent (input signal) and efferent (output signal) has the potential to create a reverberating loop of activity (Kistler & De Zeeuw, 2003). Some have suggested that reverberating loops may underlie the regularity of certain EEG waves (for example, the 40 Hz gamma wave, discussed later, which may serve to synchronise brain activity). But confining ourselves to the cerebral cortex, Szentágothai has noted the "massive reentrant circuitry" in these higher centres where efferent and afferent meet, and has wondered at the "crucial significance" this special modular architecture may have (Szentágothai, 1984). One effect of the intertwined reflex arcs between afferents and efferents might be the establishment of a "circular chain" of reciprocal connections (Szentágothai, 1979).

As shown in Figure 1, efferents and afferents combine in cerebral cortex (layers III, IV, and V), which is largely populated with pyramidal cells (pyr). The afferents are coloured blue, and the efferents red (see bottom yellow square), and the green patch is an example of where the two types touch, creating a possible reverberant loop and the potential for feedback. Feedback can happen whenever a motor and a sensor are connected and, like a microphone near a loudspeaker, the loop will create continuous oscillation if the gain is above unity.

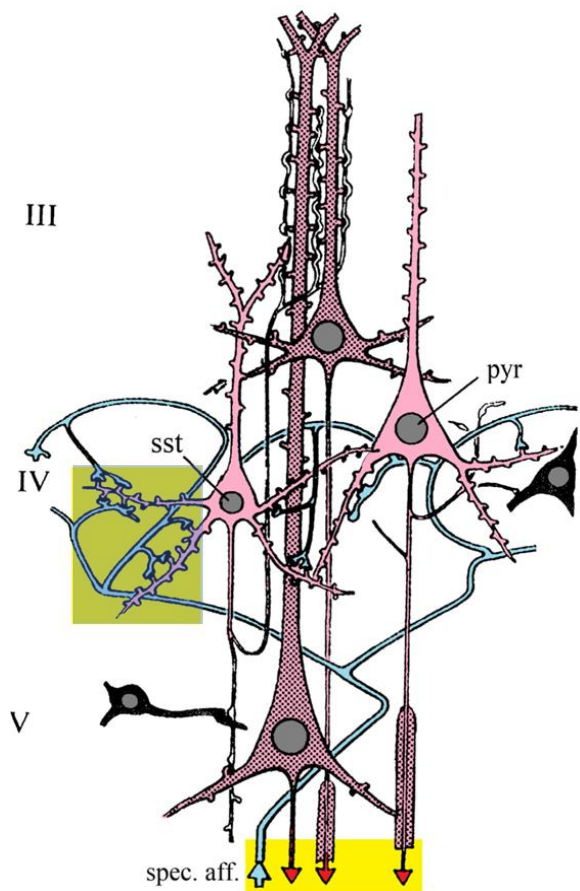

**FIGURE 1.** The anatomy of the cortical minicolumn or psychon. It is made up of pyramidal cells within layers III–V of the cerebral cortex which are interconnected so that they create feedback and a simple neuronal oscillator. Feedback occurs due to contacts between afferent (ascending or sensory) nerve fibres shown in blue, and efferent (descending or motor) nerve fibres shown in red. The two classes of nerve fibres are highlighted by the yellow rectangle. The places where sensory and motor nerves touch (e.g., in green square) lead to feedback between the two complementary types (as with a microphone near a loudspeaker) and this produces *feedback cooling*, lowering the effective temperature. (pyr = pyramidal cell; sst = spiny stellate cell; spec. aff. = specific afferent) From Szentágothai (1979) and reproduced with permission of MIT Press (all rights reserved, © 1979 Massachusetts Institute of Technology).

The result of continuous feedback is shown in **Figure 2**, where a broader view of a cortical minicolumn or psychon is shown (cortical layers I to V). This figure, taken from Eccles (1990), shows what may happen when the same pyramidal cells act cooperatively: this time the cells are surrounded by an envelope of colour, a coherent field that Eccles identified as a psychon. It is worth noting that Eccles did not speak of feedback explicitly, regarding coherence as arising from cooperative activity at synapses, but it is suggested that the end result is the same: a point where brain and mind interact and give rise to coherent quantum mechanical effects. Eccles appears to incline towards some generalised sort of coherence in the psychon, although the oscillator description given here (following Stapp) suggests that magnetic effects may be more relevant. Whatever the actual case, there is overlap between one psychon and another, potentially giving rise to long-range coherence within the cortex. The different colours in Figure 2 are meant to suggest different qualia, depending on the origin and destination of the underlying nerve pathways.

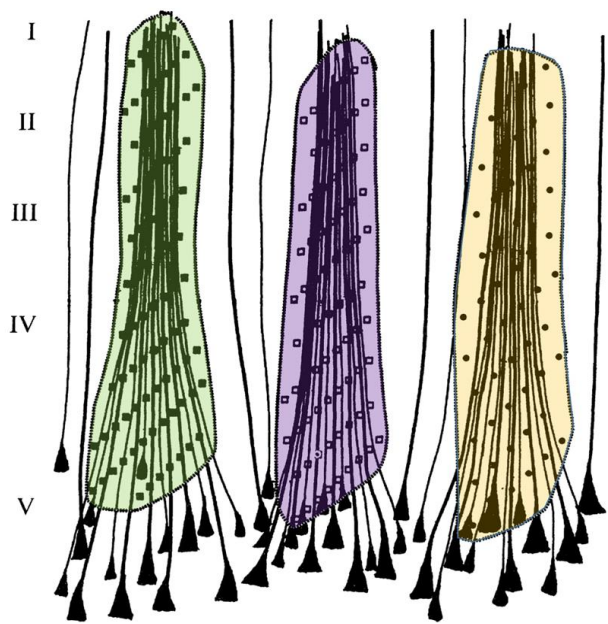

**FIGURE 2.** How macroscopic quantum coherence arises in a cortical minicolumn or psychon. This is a zoomed out view of Figure 1 and shows the fields (colour) which Eccles (1990) called psychons. The fields are here considered magnetic fields generated by electrical currents circulating within each neuronal oscillator, which act like a miniature coil. When the coloured envelopes overlap, the whole cortex becomes a unitary Bose–Einstein condensate – a macroscopic quantum mechanical phenomenon that, as speculated here, gives rise to consciousness. Modified from Eccles (1990) with permission of The Royal Society.

There are perhaps hundreds of pyramidal cells within a single psychon. For example, in a diagram by Beck (2001), his Fig. 7B shows more than 260 pyramidal cells extending over layers I to VI. Within each psychon and its bundle of dendrites, there are perhaps 100,000 boutons in which synaptic contact can take place. Eccles and Beck thought that psychons arose through a quantum mechanical process occurring at synapses, so their quantum mechanical analysis was based on the stochastics of what was happening in the synaptic cleft. Beck (2001) makes a general statement that there is "cooperation" among individual cells, but he leaves this process largely unspecified, although it could be "stochastic resonance" (p. 109).

This paper proposes that feedback loops within a minicolumn give rise to a miniature magnetic field, which, overlapping those from neighbouring minicolumns, establishes a single quantum mechanical Bose–Einstein condensate, as described later. The end result is that all the pyramidal cells within a single psychon will operate coherently. The psychon is therefore the core quantum mechanical unit, not the individual synapses of the pyramidal cells. If the fields of one psychon are quantum mechanically entangled with those of neighbouring psychons, the whole of the cortex, containing perhaps 40 million psychons, might operate as a powerful quantum computer. In other words, consciousness might be a macroscopic quantum entity extending over all the cortex, and is so entangled that it operates as a single superconducting sheet (Freeman & Vitiello, 2016).

Stapp is more inclined to see coherence as generated by the Zeno effect and its inverse. Nevertheless, both provide a common locus for where mind and brain meet: the cortical minicolumn. The new element provided by this paper is to suggest an important role for feedback cooling within the minicolumn, and this is now addressed.

## 4. Feedback cooling

The account above relies on feedback cooling to produce low effective temperatures in the cortex, and in this section we consider the phenomenon in more detail. In summary, feedback cooling is a process of applying feedback to a system so as to reduce thermal agitation, allowing quantum behaviour to emerge above thermal noise (Kawamura & Kanegae, 2016); this is the same process as used in a gravitational wave detector when a mirror is suspended in a vacuum and a laser beam and feedback is used to cancel out the mirror's thermal motion (Corbitt et al., 2007). The feedback narrows the vibrational bandwidth and this effectively cools the system. The idea was investigated in the context of animal sensory systems by Bialek, and here we give a broad account.

Bialek's investigations began with his PhD thesis in which he examined the boundaries between the macroscopic and the microscopic (Bialek, 1983). His thesis begins by calculating Schrödinger's equation for a one-ton brass bar. Due to thermal noise, it will contain phonons reverberating through it. As the bar is cooled, the number of phonons will be reduced, but even if we cool it to absolute zero there will still be spontaneous fluctuations in position due to quantum mechanical uncertainty. There will be a point where Brownian motion due to thermal energy equals quantum zero-point motion, and Bialek sought to discover where that cross-over point might be. In particular, he investigated how the same principles might apply to biological sensory organs. It turns out that the sensory organs of animals are amazingly sensitive and extend down to at least thermal limits (1 *kT* of energy, where *k* is Boltzmann's constant and *T* the temperature). He wondered if any biological system operated below the thermal limit, and considered ways by which the gap between the quantum and the classical might be bridged. It is here that he introduces the principle of feedback cooling, and he spent several years investigating how it might apply in biology. Applied to hearing, for example, he calculates that a frog can detect displacements that are smaller than $10^{-11}$ m, which is broadly comparable with quantum limits.[4] In another challenging paper, he makes the case that the faint sounds emitted by the human cochlea as otoacoustic emissions (and recorded with a sensitive microphone in the ear canal) are also at levels which approach quantum limits (Bialek & Wit, 1984).

As it happened, in the end he considered these efforts a failure and abandoned the attempt (Bialek, 2012).[5] Nevertheless, his approach provides a keen challenge: in a sensory system, can feedback cooling be used to overcome thermal noise and move us towards lower quantum limits? The principle of feedback cooling is simple but profound – that amplifiers can be used in place of refrigerators[6] – and we believe it is worth persisting with the idea, for it injects fresh life into the proposals of Eccles and Stapp. Recent work in physics has made considerable progress in using feedback cooling of nanoparticles suspended in an optical cavity. For example, Gieseler and colleagues have cooled a laser-trapped nanoparticle, 0.14 µm in diameter to 50 mK (Gieseler et al., 2012). Similarly, as cited by Bialek, it is possible to cool a suspended 1 g mirror to 7 mK (Corbitt et al., 2007). Theoretically, at least, it is evident that it is possible to cool a particle to its quantum ground state, so that the occupation number is less than one quantum (Manikandan & Qvarfort, 2023). Some work has entertained the idea of levitating a living tardigrade (Romero-Isart et al., 2010).

---

[4] Bialek calculates that at absolute zero, quantum noise is larger than $10^{-12}$ m. At room temperature, noise displacements are some 40 dB greater than this. He concludes that the ear is a macroscopic quantum detector (Bialek 1983a, p. 50) and that stereocilia must possess a filter with a bandwidth of 50 Hz or less. As to the amplifiers used, he concludes that a necessary quantum limited amplifier must have a long memory for quantum mechanical phase – they must be *coherent* (op. cit., p. 70).

[5] "… (confession time) there was a time when I worked very hard to convince myself that these quantum limits to measurement could be relevant to biological systems. This project failed, and I would rather not revisit old failures, so let's skip this one." (loc. cit., p. 237).

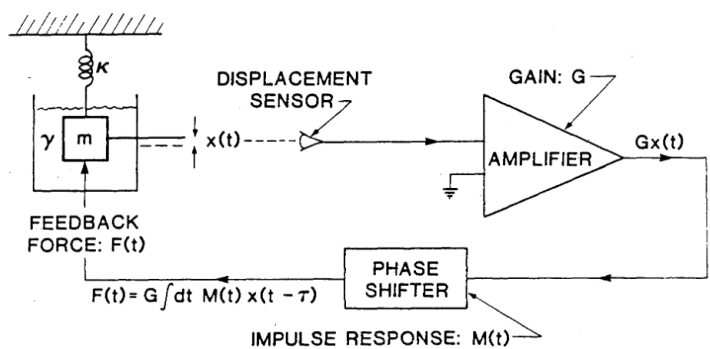

**FIGURE 3.** How the effective temperature of a mass hanging on a spring and surrounded by liquid can be reduced by using a sensor and applying feedback to counteract vibration. Effectively, the circuit acts as a refrigerator. Initially, the mass experiences drag $\gamma$ but feedback provides additional drag $\eta$, so that the total drag becomes $\gamma + \eta$ and the effective temperature, $T_{eff}$, becomes $T\gamma/(\gamma + \eta)$, where $T$ is the initial temperature. The sensor could be a piezoelectric material; similarly, the feedback force could also be supplied using a piezo material. From Bialek (1983) and used with permission.

How does feedback cooling work? **Figure 3** is a diagram of a mass hanging on a spring, and by adding a displacement sensor and feeding back a signal so as to cancel perceived motion, the centre-of-mass system can be effectively frozen. That is, the feedback reduces the mass's thermal vibration so that its bandwidth is less and its effective temperature much lower. As Bialek explains, the reduction of noise via feedback has no limit, since it is possible to narrow the bandwidth as much as we like, although it is paid for in terms of the response time of the detector and the power expended in pushing on the mass. So far as the mass is concerned, it experiences the feedback force as if it were extra drag, and in this way the effective temperature, $T_{\text{eff}}$, can be indefinitely reduced (see caption). In a similar way, the nitrogen atoms in the Quantum Brilliance device are confined by the crystal lattice of the diamond, allowing them to have long-lived quantum states.

---

[6] Bialek (1983), p. 43.

The key message is that through sensing an object and applying negative feedback, it is possible to create a system that is effectively colder.[7] In the same way, it is suggested that an arrangement of feedback circuits in the pyramidal neurons of the minicolumns could reduce their effective temperature to a level where thermal effects are diminished and quantum mechanical processes come into play. In particular, cooling may allow a Bose–Einstein condensate to form in which bosons crowd into low-energy states and, via quantum mechanical correlations, the material takes on long-range order. In this way, the cortex effectively becomes a superconductor. Another way of seeing what is going on is to recognise that when a material is cooled its atoms begin to take on precisely defined momenta (nearly zero), which means, quantum mechanically, that the complementary variable (position) must be highly indeterminate (via the Heisenberg uncertainty principle). Cooling thus produces a fundamental change of state, a condensation, in which activity in one part of the brain is quantum mechanically linked to all the other parts.

Another thing to note is that in a typical superconductor only one entity in 10,000 participates in the phase change, even though it leads to a dramatic change in macroscopic behaviour. As this small fraction enters the lowest momentum state, the complementary variable, position, becomes increasingly ill-defined until eventually there is shared identity and long-range order. The message here is that cooling does not have to be total: it is enough if the effective temperature of the minicolumns is reduced and only a fraction of the quantum states are low-momentum states (the states with centre-of-mass coordinates reflecting low effective temperatures), even though the bulk of the material remains at room temperature and most of the molecules (or nuclei) occupy high momentum states.

---

[7] Its *centre of mass* temperature is low, even though its *bulk temperature* remains near room temperature. Nevertheless, effective cooling allows quantum phenomena – occupation of low-level quantum states – to be observed (Romero-Isart et al. 2010), and this provides a route for Bose–Einstein condensation.

Another helpful perspective on Bose–Einstein condensation comes from Marshall (1989) where he argues that consciousness may arise from a system with long-range order in which the constituent quantum mechanical bosons have overlapping wave functions, a picture consistent with the minicolumn model. Marshall lists how a Bose–Einstein condensate fulfils the requirements for a substrate of consciousness: it is extended in space, capable of myriad states, and cannot be separated into parts with individual identities (ibid., p 79). He notes that Bose–Einstein condensates normally exist only at extremely low temperatures (such as in superfluids and superconductors), although it is theoretically possible for a pumped phonon system (such as that suggested by Frölich) to operate at room temperature. Marshall therefore favours the pumped phonon system, as does Penrose (2022), but here we suggest that the low-temperature approach provided by feedback cooling is more realistic.

There is one other key factor in this scheme which needs to be considered, and that is to indicate how feedback might be sensed and applied. In this connection, it is pointed out that the material of which nerve axons are made has the unique and unusual property of being piezoelectric (Costa et al., 2018; Tasaki, 1999; Tasaki & Iwasa, 1982). What this means is that when an action potential passes along an axon, its diameter momentarily changes.[8] Most piezoelectric materials behave reciprocally, meaning that a change in diameter is also accompanied by a change in voltage – see Ludwig et al. (2001) for the biological example of prestin, a bidirectional piezoelectric material in cochlear outer hair cells. Also highly relevant are reports that dendritic spines in pyramidal cells mechanically twitch when electrically stimulated (for context see Crick (1982)). Thus, by using piezoelectricity and feedback, it appears possible to devise a scheme whereby a particular site within a minicolumn could be effectively frozen. Any change in diameter due to thermal noise, for example, could be compensated for by a change in voltage.[9] The piezoelectric

---

[8] There has been a long history of similar observations, their significance has not been well appreciated. Liberman and colleagues (Berestovskii et al. 1970) saw changes in the birefringence of nerve fibres during passage of an action potential and connected that to transitory structural changes.

[9] It is again helpful to consider the effects of ephaptic (electrical) potentials at gap junctions. In this way the more complicated effects of action potentials in the synaptic cleft can be treated separately. For a perspective on analog and digital processing in neural networks, see Debanne et al. (2013), who point out that

property of nerves deserves greater attention, as it may help explain why this special property has arisen. As **Figure 3** illustrates, piezoelectric properties are required in order to make feedback cooling work.

## 5. Quantum accelerators

When quantum computers are mentioned, most people think of a large room filled with rows of cryogenic equipment and racks of electronics.[10] A team based in Canberra and Stuttgart are changing that conception. They have developed a small room-temperature quantum computer, small enough to hold in your hand, that they describe as a "quantum accelerator" (Doherty, 2021). Doherty and his team are part of a research spin-off company who have developed, and now offering for sale, a desk-top device that plugs in to a conventional computer and speeds up difficult computations (such as cryptographic problems), greatly increasing computing power. The quantum accelerator developed by the Australian–German company Quantum Brilliance[11] is a solid-state device based on atoms of nitrogen embedded within a diamond lattice and contains a total of 5 quantum bits (qubits), although the aim is to increase that number considerably.

The core idea of the present paper is that the human brain also contains a quantum accelerator which speeds up normal neural processing. The brain, we suggest, can either motor along using its standard neural networks to deal with inputs and outputs, or it can add on a quantum accelerator to perform faster and more complex computations. (The speed advantages of quantum computation were first laid out by Deutsch and Jozsa (1992)). In the first case, the processing is done unconsciously; in the latter case, the quantum processing involves qualia. The owner of the computer, the self, is vividly aware of these qualia but not of the subconscious activity. That distinction will be brought out in section 8, where the peculiar state of sleepwalking is discussed.

Room-temperature diamond computing was discovered in 2001, and work since then has demonstrated that it works. The practical problems are fabricating them

---

the effects of action potentials and digital signalling have been overemphasised at the expense of electrical synapses and analog signalling.
  [10] For reviews of quantum computers see Ladd et al. (2010) and de Leon et al. (2021). An engaging introduction to the topic is Feynman (1986). Perspectives on theories of mind can be found in Aaronson (2013).
  [11] https://quantumbrilliance.com.

consistently and interfacing them with conventional computers, and Quantum Brilliance is working on these aspects.

The quantum accelerator consists of an array of nodes within a diamond crystal in which each node contains a nitrogen vacancy (where a nitrogen atom substitutes for a carbon atom). These NV defects contain a cluster of nuclear spins – the intrinsic nitrogen nuclear spin and a handful of nearby $^{13}$C nuclear spin impurities – which act as the qubits of the system. The NV centres act as quantum buses that initialise and readout the qubits, and quantum computations are controlled by microwave, optical, and magnetic fields (which is proprietary information). The remarkable part of the system is the long electron spin coherence time, about 1 ms, which is the longest of any known solid-state electron at room temperature. This allows the NV centre to initialise the spins, processing to happen, and read-out to take place without decoherence setting in.

Development is proceeding, with the short-term goal being to build quantum accelerators with more than 50 qubits, which will outperform CPUs of comparable size and weight (computing power increases exponentially as the number of qubits). Ultimately, there will be massively parallel quantum accelerators which will far exceed the capabilities of today's classical supercomputers. Applied to neural networks, it has been calculated that a quantum algorithm will be $n^5$ times faster than the best classical algorithm (Doherty, 2021, p. 78). The conclusion to be drawn here is that nature would be able to create immense computing power of unrivalled sophistication by drawing on quantum computation, especially if *n* were large. A discussion of how this enormously expanded capacity might be a powerful factor in evolution is provided later.

Decoherence is the biggest enemy of a quantum computer, and in making a better quantum computer the goal is always to minimise it (Schlosshauer, 2019; Stamp, 2006). The field is highly technical, but there are certain configurations of quantum systems – for example, *decoherence-free subspaces* – where quantum information can be encoded such that it is immune from environmental intrusion (Schlosshauer, 2019, Sec. 5.1).[12] See Lidar

---

[12] For the human brain, we are probably looking for decoherence times in the millisecond range, even though, at room temperature, the decoherence time has been calculated as about $10^{-20}$ s (Tegmark, 2000). This means that ways to create decoherence-free subspaces are important. One possible method outlined by Schlosshauer is to create environmental symmetry so that each qubit operator couples to the environment in exactly the same way. If so, in a very large qubit system "essentially every state in the system's Hilbert space

(2014) for additional details of how to minimise decoherence and Fortunato et al. (2003) for an actual example. Speculations of the link between consciousness and the existence of some long-lived spin states in phosphorus atoms have already been made by Fisher (2015).

It is notable that recent investigations have used quantum techniques to perform noninvasive thermometry on living cells (Kucsko et al., 2013; Silletta et al., 2019), and this is a potential avenue for experimentally detecting localised cooling effects in the cortex. The Kucsko work is based on NV centres in diamond, and it is able to detect temperature changes within a living cell of less than 2 mK. The Siletta group's approach involves water and sodium ions and can achieve accuracy of a few tenths of a degree. It may ultimately be possible to use nuclear magnetic spectroscopy (NMS) to measure temperatures within a conscious brain: if an NMS probe revealed a narrow line-width, this would indicate that the signal has encountered low-temperature nuclei (Buonocore & Maddock, 2015).[13] In principle, such non-invasive experiments are relatively simple, but technically challenging.

## 6. Thalamocortical 40 Hz loops

We have suggested there is a vital functional role for feedback loops within the psychons of the cortical sheet. To elaborate a little, it is worth underlining the importance of feedback circuits within the brain, especially the 40 Hz thalamocortical loop, which is known to play a major role in regulating consciousness (Llinas & Paré, 1991; Ward, 2011). All sensory messages (except smell) reach the cerebral cortex though the thalamus. Moreover, pyramidal cells in layer VI project back to the area of the thalamus where their specific input arises, although the number of corticothalamic fibres is about 10 times larger than the number of thalamocortical fibres (ibid, p. 525). The neurophysiological literature describes how brains sustain multiple feedback circuits, often called reentrant or reverberant circuits, of many kinds, and their combined activity is thought to underlie the electroencephalogram (the EEG). The 40 Hz activity falls within the gamma band of the EEG, and a key property is

---

will be immune to decoherence" (Schlosshauer 2019, p. 29). The speculation put forward here is that coherent electrical or magnetic fields in the brain (as evident in the EEG and MEG) might affect all the qubits symmetrically and make them immune from decoherence. It is also significant that decoherence falls particularly rapidly in insulators (Stamp 2006, p. 480), and we note that the myelin surrounding a nerve fibre is a high-resistance insulator.

[13] To be clear, it is the nuclei that are cold; most of the brain remains at room temperature. This is like in a diamond-based quantum accelerator, where the atoms are at room temperature but the nuclei are effectively at 1 mK.

that it acts in synchrony over the entire cortical mantle (Llinas & Paré, 1991, p. 527–9), and this has important quantum mechanical implications. A good overview of reentrant feedback loops and how they relate to the generation of qualia is given by Orpwood (2013).

Orpwood's survey recognises the immense amount of research that has been done in the area. Although Llinas and Paré (1991) explicitly hypothesised that the thalamocortical system is "the consciousness-generating apparatus", the most influential work has been that of Edelman who has set out descriptions of how 40 Hz circuits could be the locus of conscious percepts (Edelman, 1992). Feedback can operate at different levels, but Edelman claims that the 40 Hz thalamocortical loop is fundamental and in fact he identifies it with consciousness itself (Edelman, 1992; Edelman & Gally, 2013). He considers that reverberating activity between the thalamus and pyramidal cells in the cortex could be the actual origin of qualia. Whereas that might be true, Edelman does not explain how qualia necessarily arise from what is, fundamentally, neural activity. He does call the qualia assumption "tricky", and despite a lengthy discussion, the explanatory gap, in the end, lingers.

In the model proposed here, the 40 Hz thalomocortical loop has the important role of switching cortical cooling on (awake) and off (asleep). This means the loop is *essential* for consciousness, but by itself it is not *sufficient*. So even though Llinas & Paré and Edelman both provide valuable context for how consciousness is generated, the picture remains quite general, whereas, by comparison, Eccles and Stapp pin-point the locus to specific clusters of pyramidal cells in the cortex. Although psychons are actually switched on by the 40 Hz thalamo-cortical loops, it is the tight feedback loops within the psychons themselves – a very local loop of roughly 1 kHz operating within a domain of just 60 × 1500 μm – which lead to feedback cooling and consciousness.

## 7. Evolution of consciousness

The foregoing gives insights into why and how consciousness evolved. Primates may well have evolved to the point of acting something like a neural computer (Cottam & Vounckx, 2022), but by adding feedback cooling to the same neuronal architecture, the possibility of having a quantum computer with superior processing power becomes possible.[14] Feedback cooling involves adding feedback circuits on top of the neuronal processor, and with this quantum leap (so to speak), consciousness emerged.[15] With the light of consciousness, we are able to instantly survey all our neuronal inputs and outputs, an enormous advantage in terms of decision-making and survival (Cleeremans & Tallon-Baudry, 2022). This is the adaptive significance of consciousness.

As suggested at the beginning, there is a seamless transition from unconscious neural computer to conscious quantum computer because they both share the same basic hardware – the cortical minicolumn. When thalamocortical circuits are inactive, the minicolumns act individually and unconsciously, but when switched on the units act together – their wave functions overlap – to form a macroscopically coherent, and awake, cortical sheet. The resulting conscious 'I' encompasses all of those units and can simultaneously access all their collective information.

The paper by Marshall (1989) describes how consciousness arises from a Bose–Einstein condensate, work which provides a nice unifying account of how quantum mechanics underlies consciousness. When Marshall asks why consciousness exists in the world, his answer is close to the one outlined here: he says that "consciousness is a necessary concomitant of a certain kind of holistic system which, if it were intervened between stimulus and response, would enable the response to be more integrated" (and hence favouring survival) (Marshall, 1989, p. 81).

Moreover, with the ability for the conscious computer to program the subconscious computer, one can understand that an enormous increase in adaptability and processing power becomes possible (as Searle (1992) surmised). Humans can teach themselves how to perform certain desired actions – play a sonata, perform a double back-flip with twist, even strive to understand themselves and the world around them. Consciousness was a great advance, and it can be reasonably argued that many animals have also learned to use it. A not unlikely proposition is that all animals that sleep – most of them – might be conscious beings (Descartes and his followers notwithstanding, see Huxley (1874)). For the other side of consciousness is that it requires expenditure of energy – feedback circuits need energy to operate – and so all conscious creatures need restorative sleep.[16] Sleep is the time when all conscious creatures become unconscious, and during that phase all the reprogramming of the physical brain takes place and the energy circuits are replenished. Of course, some level of consciousness remains during dreaming or REM sleep, but during NREM sleep, when sleepwalking occurs, everything occurs "in the dark," when automatic subconscious circuits take over.

---

[14] According to Cottam & Vounckx (2022), an animal's computational ability is a powerful evolutionary driver, and so quantum computation is a natural avenue for advancement. They point to the fine axonite mesh, free of action potentials, as the locus for this.

[15] Godfrey-Smith (2016) inclines to the view that consciousness is not an on/off affair, but that it is graded (p. 87). See also Godfrey-Smith (2021), p. 262. However, if consciousness is the result of a phase transition (a Bose–Einstein condensation), it cannot come into being gradually; it either exists or it doesn't.

A number of philosophers have been puzzled by why evolution should have taken a path from unconscious reflexes to complex conscious behaviour, in humans at least. The puzzle is clearly set out by Lockwood (1998), where he notes that sentience (consciousness), which "sticks out like a sore thumb" (ibid., p. 84), must confer some adaptive advantage. Specifically, he assumes that there are some cognitive tasks which either cannot be performed at all without sentience, or more to the point, as put forward in this paper, that they can be performed more efficiently with sentience (Lockwood, 1998, p. 83). The parallel he makes with blind-sight is particularly apt.

Flanagan (1995), whose insights into sleepwalkers and zombies we will look at below, raises a related question, why did creatures need to be more than just "informationally sensitive"? He reports he found a total absence of credible theories (Flanagan, 1995b, p. 319). However, he does provide one clue by suggesting that in times past there might have arisen a "computational bottleneck", which led to a problem of information overload. He mentions serial computers and parallel computers, and this appears to be on the right track, although he mistakenly, I think, identifies consciousness with the serial variety. The real distinction, I suggest, should be that between a classical computer and a quantum computer. The standard computer is no match for the quantum computer, which can operate in parallel and solve wholly different classes of problems (Deutsch & Jozsa, 1992; Penrose, 1989; Schlosshauer, 2019). Moreover, a quantum computer can be built *on top of* a classical computer (as a quantum accelerator), enabling a sentient being to evolve and overcome computational bottlenecks. All it requires is for a mechanism to be devised to cool down the dendritic cluster in the cortex and shift it into quantum mode. With quantum processing enabled, the being becomes fully conscious and its skill repertoire speeded up and refined. The difference in performance, we suggest, is like that between a sleepwalker and the same person fully awake. It is now possible to understand why such an intelligence-boosted creature would win an evolutionary battle against what Flanagan calls "zombie-like information-sensitive organisms" (Flanagan, 1995b, p. 321).

---

[16] It is known that all mammals undergo REM sleep, which could be a marker that these animals, at least, have dedicated consciousness-raising circuits in which they dream (Flanagan, 1995a). Octopuses are also thought to dream in that they undergo periods of rapid skin colour change while asleep (Godrey-Smith, 2016).

Thomas Nagel is similarly puzzled when he looks at evolutionary reductionism (Nagel, 2012). If materialism cannot explain the fundamental mind–body problem, it certainly can't provide a credible account of how conscious minds evolved from unconscious ones, however hard it tries (Chs. 1–4, and the list of references in their endnotes). It will take more than "just the lacing of life with a tincture of qualia" (p. 44) and at some point he thinks evolutionary biology will need to make a fresh start, starting with an understanding of the mind and the important role it plays in living creatures. The basic question is how can a single accidental mutation create a conscious creature from one that has none, and why should that mutation have more evolutionary success than its unconscious rival? Nagel finds that, despite its great achievements, the prevailing austere doctrine – that everything is due to the action of forces on physical matter – is untenable. That can only be part of the truth, he believes, and suggests the answer must lie in some sort of neutral monism, rather than dualism. As suggested earlier, if the missing ingredient is quantum mechanics, then dualistic interactionism, as favoured by both Eccles and Stapp, may also be able to overcome this sticking point. Interestingly, Nagel mentions concepts such as relativity, but quantum mechanics itself is never on the table. He accepts that mind is a biological phenomenon (p. 45), but, as matters stand, he cannot see how to explain it naturalistically, and regards the resources of physical science as inadequate.

This paper proposes that the explanation of how the physical and the mental sides of an organism can develop together (Nagel, 2012, pp. 46-47) lies in the realm of quantum computers. The co-existence in one brain of both a classical and a quantum computer renders the facts intelligible. The emergence of a quantum computer as an add-on to an existing neural network computer explains why, in Nagel's words, "the appearance of complex organisms, and not merely behaviourally complex organisms, was likely" (p. 48). The quantum computer allows the physical and the mental to work together to achieve a wondrous thing – the universe waking up and becoming aware of itself (p. 85).

## 8. Of sleepwalkers

Reflecting on our own consciousness brings us face to face with the notoriously 'hard' questions of consciousness (Chalmers, 1995, 1996). Why is our mind so immediately present and yet so elusive? Here we will follow the approach taken by Owen Flanagan based on reflections about sleep, dreams, zombies, and sleepwalkers (Flanagan, 1995b), which seems to offer an insightful perspective into the factors which separate the conscious from the unconscious.

Stapp (1993) recognised that sleepwalking is a distinct, unconscious state. When defining consciousness (p. 234) he notes that it "is extinguished during dreamless sleep … [and is] … absent in the state of somnambulism." Penfield (1975) discussed the similar behaviour of patients undergoing petit mal seizures who are "totally unconscious", even though they are capable of continuing well-rehearsed patterns of activity, such as driving or playing the piano. Searle (1992) was struck by these cases and wondered why all behaviour couldn't be unconscious, and his answer was that consciousness provided extra "flexibility and creativity" (p. 108).

As set out by Bell (2024), the distinctive feature of sleepwalking is that body movements are slow. Not only is walking slow, but all actions appear to be slow and clumsy. Speech, for example, is slow and indistinct. All these things are just the sort of symptoms expected if the processing power of a computer controlling a robot were suddenly compromised. The frame rate of its display would immediately plunge, and it is suggested that this is similar to what might happen if a computer were to suddenly lose its accelerator. Devoid of consciousness, and relying only on its pre-programmed brain circuits, the sleepwalker must necessarily act slowly. It is no wonder that the eyes are glazed and the face expressionless because the somnambulist is in deep sleep. We are reminded of Lady Macbeth of whom it was said: "her eyes are open," but "…aye, their sense is shut." In philosophical terms, there is *nothing it is like* to be a sleepwalker, or more commonly, no feeling of *here-I-am*.

It is noteworthy that, electrophysiologically, the prefrontal cortex and hippocampus of sleepwalkers are profoundly disengaged, which reflects inactive memory circuits and automatic behaviour (Krouse, 2022). At the same time, the distinctive 40 Hz oscillation between thalamus and cortex, a marker of qualia when one is awake or in REM sleep, is absent (Flanagan, 1995a, 1996; Llinas & Paré, 1991).

On this basis, we make the claim that sleepwalking is a case of a slow neural network computer operating unconsciously, but when the thalamocortical loops are active and feedback cooling is engaged, the person wakes up. We will not enter into the details here, but a companion paper (Bell, 2024) explores the matter in more depth.

## 9. Qualia and some metaphysics

As Chalmers has forcefully reminded us, quantum mechanics by itself does not solve the hard problem (Chalmers, 1996). There is still an explanatory gap between the states of bosons and the subjective impression of a luminous patch of blue; between first-person experience and third-person fact (Schurger & Graziano, 2022). As Lockwood (1998) puts it, why do certain physical events in my brain give rise to subjective correlates when agitation of molecules in a stirred cup of tea presumably does not (ibid., p. 85). Lockwood inclines to the view that it is just a brute fact. We are entering metaphysical waters here, but to keep matters brief we concentrate on two clear landmarks.

First, there is the commendably readable paper which considers the metaphysics of quantum mechanics and clearly maps out the territory (Marshall, 1989). Marshall begins by describing three realms – the mental, the bodily, and the quantum. Then, from the existence of two fundamental properties of consciousness – one its unity even though extended in space, and two its complexity – he concludes that *the brain processes corresponding to states of consciousness cannot be described by classical physics* (ibid., p. 74). That is, the mental and bodily realms both emerge from different sorts of underlying quantum stuff (as he illustrates in his Figure 2). He therefore suggests that neutral monism, or perhaps some sort of panpsychism, might be an appropriate way of describing the world, although he admits that dualists may well be able to live with the quantum mechanical picture.[17] Neutral (or dual-aspect) monism is a position that is compatible with the "general monism / panpsychism" advocated by Nagel (2012, p. 56,57), the panpsychism of Strawson (2019) and Stubenberg (1998), and the protoconsciousness favoured by Chalmers (1996) and Penrose (2022). Comprehensive discussions of the philosophical options are presented in Robinson (2018) and Mørch (2023).[18] If the whole world is quantum mechanical through and through then classical physics actually becomes an unwanted hindrance. A wide-ranging review of how quantum mechanics relates to classical physics can be found in Schlosshauer (2019). According to him, the key is the notion of decoherence, which acts as the boundary between the quantum and classical realms (ibid., p.2). As remarked, Nagel never considers quantum mechanics in his work.[19]

In this context I raise a second point. As a speculation, it might be that qualia need to exist because we have two different coordinate systems in our brain, each associated with the two computers residing there. We have the position – the cartesian coordinates – of the pyramidal cells within our cortex, and the coordinates of the Hilbert space in which quantum processing is done.[20] The suggestion is that these two systems are separate but need to be kept in alignment, and so qualia are necessary to "label" the results of the quantum accelerator, distinct from those of the basic reflex computer.

To expand this a little, the low-level computer is used to calculate reflex reactions. It operates unconsciously and is involved in sleepwalking. Its coordinates are effectively those of the pyramidal cells, and it provides a map, a body image, of neural inputs and outputs. Here we point to the octopus, which has separate reflex computers in each of its eight arms and a second high-level computer in its brain, but both need to be able to refer to the same things. Likewise, we have a neural computer formed from pyramidal clusters and a second quantum computer – the accelerator – sitting on top. The neural computer operates in cartesian space, and the accelerator operates in Hilbert space, but the two need to have a common reference point. This means there needs to be strict mapping between the computers – as the octopus case illustrates[21] – so that both coordinate systems operate seamlessly. The suggestion made here is that qualia, which are the eigenstates of the quantum computer, need to be clearly distinguished as products of the mental self but still refer back to the same (spatially defined) inputs and outputs. In brief, qualia may be the quantum mechanical 'labels' that allow mind and body to synchronise.[22]

---

[17] I take this to mean "pure" dualists, and not interaction dualists such as Eccles and Popper who are willing to embrace quantum mechanics. So both Marshall and I agree that consciousness does have an "isomorph" in physics, as he puts it (Marshall, 1989, p. 81).

[18] Even though physicalism has become the default position in theories of consciousness, one potentially attractive position for physicists wanting to bridge the epistemic gap (i.e., acknowledging that consciousness is more than the purely physical) might be dual-aspect monism (Mørch 2023, Ch. 4). Here it is worth noting how macroscopic quantum states, like superconductivity, can avoid "the combination problem" – that is, how a single unified experience could emerge from the fusion of multiple protoconscious particles. In superconductivity there is fusion of particles via Bose–Einstein condensation, and it seems that feedback cooling might allow such fusion to occur in the brain. Mørch (2023) notes: "If genuinely novel physical properties or behaviors arise in macrophysical systems, this could be explained by these systems having a novel, fused intrinsic nature" (p. 66). As she expresses it in her earlier PhD thesis, "science constrains but underdetermines mental structure." (Mørch 2014, p. 191).

[19] Nagel points to Sorell, Strawson, Hartshorne, and Whitehead as some of those supporting his position (his footnote #16). In another footnote (#11) he delivers a cutting attack on reductive materialism, and on p. 26 he supplies apt quotes from Schrödinger, de Broglie, Eddington, Planck, and Einstein to reinforce his argument that mind and matter are different aspects of the same stuff.

[20] Schlosshauer (2019) explains that "Hilbert space is a vast and seemingly egalitarian place. If $|\psi_1\rangle$ and $|\psi_2\rangle$ represent two possible states of a quantum system, then quantum mechanics postulates that an arbitrary superposition $\alpha|\psi_1\rangle + \beta|\psi_2\rangle$ constitutes another possible physical state" (ibid., p. 2).

## 10. General discussion

We have set out a speculative, but considered proposal for how the human brain may operate as a room temperature quantum computer. The model is fully specified anatomically, and it offers explanations for how and why consciousness evolved. At its core, the picture is one in which feedback cooling in the cortex allows macroscopically coherent quantum mechanical states to be established in assemblies of pyramidal cells, and which, when energised, condense to a macroscopic quantum state, as in a superconductor, that can be identified with the phenomenal mind. In brief, one could say that, quantum mechanically speaking, mind is matter waking up.

The feedback cooling model places emphasis on the minicolumn as the core unit, rather than on individual neurons, which have been the focus of much earlier work (e.g., Fields et al. (2022)). The modelling of Fields and colleagues called on each neuron to have quantum mechanical coherence in order for it to become a computational unit, but the acknowledged problem is how to shield it from thermal decoherence. Various schemes have been proposed such as Fröhlich pumping and shielded states (Davies, 2004; Fields & Levin, 2021), but these remain speculative. Here, the focus remains on the minicolumn – an assembly of 100–200 neurons arranged as an oscillator – which provides a ready scheme for creating positive feedback and therefore effective cooling. The need for a biological system to operate at temperatures near absolute zero if it is to process information quantum mechanically has been set out earlier (Matsuno & Paton, 2000) and later emphasised by (Igamberdiev, 2004, 2007). Interestingly, the recent Fields paper (Fields et al., 2022) in fact devotes a section to minicolumns and concludes that they probably act as functional units called Markov blankets, but this is a technical issue beyond our present scope.

---

[21] See Godfrey-Smith (2016).
[22] Any mismatch between the two computers could give rise to illusions – either minor and temporary (such as perceiving stationary circles as rotating) or major and long-lasting (such as feeling pain in a phantom limb). It would be of interest to test whether sleepwalkers, who operate only with the lower-level computer, are susceptible to common visual or auditory illusions.

Perhaps an apt conceptual model of what we have been describing is Indra's net, sometimes called the pearls of Indra, which in Buddhist tradition has been likened to the mind. Each pearl is perfectly reflective, and reflects the images of all its neighbouring pearls, just like each cortical column is reciprocally connected to all its neighbours. The resulting macroscopically coherent quantum mechanical state gives rise to a powerful quantum computer that we identify with consciousness. As noted in section 8 on sleepwalkers, the 40 Hz loop is highly active during periods of wakefulness and REM sleep, which is just when consciousness arises. At other times, the loop is quiescent and we are unconscious, even if sleepwalking. Ultimately, then, our thalamus is responsible for waking us up (Redinbaugh et al., 2020; Ward, 2011).

There are a numerous quantum mechanical models (Atmanspacher, 2020; Gao, 2022; Pylkkanen, 2018), three of the major proposals being OrchOR (Hameroff & Penrose, 2014; Penrose, 2022), Tononi's Integrated Information Theory (Oizumi et al., 2014), and the quantum field theory of Freeman and Vitiello (2016). A useful context for making side-by-side comparisons is provided by Hviid Del Pin et al. (2021), although there is not the space here to delve further. The theory of Hameroff and Penrose speculates that consciousness arises due to the orchestrated objective reduction, by gravity, of quantum states occurring in assemblies of tubulin molecules in the brain. As we have mentioned, however, a long-standing criticism (originally raised by Tegmark) is that thermal noise is so strong, and gravity so weak, that decoherence effects will overwhelm any tendency towards quantum coherence (for a good survey, see Rosa and Faber (2004)).

We make the broad claim that the feedback cooling model seems to avoid a number of problems that have been associated with established models of consciousness (Doerig et al., 2020). In particular, it appears to avoid specific criticisms that have been levelled at the OrchOR model (Grush & Churchland, 1995) and also avoids some of the intricacies and arbitrariness of Tononi's integrated information theory (Oizumi et al., 2014). We claim that the distinguishing feature of consciousness is not that there is some simple numerical measure – which misses the point about the puzzle of qualia – but whether subjects are able to sense they are conscious – here I am – and later remember and report it. This of course calls for a detailed treatment of memory and forgetfulness, which is beyond what is possible here.

We have already mentioned Tegmark's calculations (Tegmark, 2000a, 2000b, 2014) of the rapidity of decoherence in the brain. Litt (2006) raised a similar question-mark when criticising the Hameroff and Penrose model. The crucial point is that all these calculations assume room temperature conditions. In this context, it is helpful to look more closely at the calculations of Rosa and Faber (2004), which specifically address the effect of temperature on quantum state lifetimes. Their Table 1 shows that, at ambient temperatures, decoherence times-scales are orders of magnitude shorter than superposition time-scales. Importantly, they do consider the case of very low temperatures, and here they find that decoherence times become comparable with superposition times. The possible conclusion is that feedback cooling appears to be a promising way for the brain to establish and sustain coherence, and by switching cooling on and off consciousness can also be turned on (awake) or off (asleep). With feedback cooling active, macroscopically coherent quantum states can emerge. The direct parallel is with superconductivity, where cooling is necessary in order for macroscopic quantum coherence to emerge.

According to a rebuttal of Tegmark by Hagan et al. (2002), it may be possible for the decoherence time at room temperatures to be as long as $10^{-4}$ s, in which case it is unclear whether cooling is absolutely necessary for consciousness to appear. Nevertheless, lower temperatures will always make coherent states easier to achieve, as set out in Hameroff (2007). Of particular interest, the title of Hameroff's 2007 work is "The brain is both neurocomputer and quantum computer", giving rise to an expectation that an explanation might be given of how the two computers are related. Unfortunately, the relationship is not disclosed, the closest approach being the statement "[d]uring general anesthesia in the absence of consciousness, neurocomputation in the brain continues" (p. 1041). Elsewhere, in Hameroff (2010), it is explained how a "conscious pilot" assumes control of an otherwise non-conscious "auto-pilot neurocomputation", but again the relationship is unclear. Why are microtubules sometimes generating consciousness but at other times not? Why don't microtubules forever keep us conscious?

Despite the complexities of the OrchOR model, the 2010 paper falls short of giving a definite answer, but it does point to a role for 40 Hz gamma synchrony, generated by ephaptic "sideways" connections between neurons. By way of contrast, the virtue of the feedback cooling model is that it unifies these two aspects by explaining that gamma activity is generated by thalamocortical loops, i.e., control signals from the thalamus which are explicitly wired to switch on cortical cooling in the cortex. In this context, it is significant that, according to Newman and Baars (1993), the "cerebral cortex and thalamus are, in a very real sense, 'mirror images' of each other," and connected by an extensive fan of millions of reciprocal fibres. Ward (2011) expresses it in terms of the thalamus being "a miniature map" of the cortex. That is, there appears to be a thalamocortical switch which is specifically organised to wake us up, region by region (Alkire et al., 2000; Alkire & Miller, 2005; Llinas & Paré, 1991). That being said, the role of the thalamus in consciousness remains controversial (Koch et al., 2016).

In the end, of course, the feedback cooling idea is not incompatible with the Orch OR model, in that cooling may indeed support the appearance of long-lived quantum states in microtubules. Similarly, IIT measures are very likely to correlate with the appearance of Bose–Einstein condensates, and that cooling will assist in generating conditions suitable for establishing a single superconducting sheet – a macroscopically coherent quantum state – extending over all the cortex (Freeman & Vitiello, 2016).

We make the general claim that the minicolumn model, supported by feedback cooling, provides an anatomically and physically well-specified model of consciousness, with a number of advantages. In particular, it provides a distinct neurophysiological locus – the cortical minicolumn – which can operate both classically and unconsciously (the neural computer) or quantum mechanically and consciously (a quantum computer), depending on whether or not feedback cooling is switched on. The switching between the two states matches what is known about conscious states being present when 40 Hz thalamocortical loops are active.

Of course, as Chalmers has emphasised, quantum mechanics does not solve the hard problem, and so some sort of explanatory gap inevitably remains. Why exactly should a macroscopic quantum state generate qualia? We have made a suggestion in terms of linking Hilbert space and neural space, but it will require much more psychophysical detail to make this idea explicit. Some NMR experiments on conscious volunteers may prove useful.

On the negative side, the model has limitations. A major drawback is that feedback cooling is a delicate process which has yet to be observed in a biological system. Cooling can certainly be achieved using lasers and carefully arranged systems of sensors and actuators. However, as noted previously, Bialek devoted years to investigating how feedback cooling might be applied to human sensory systems, and eventually gave up on the enterprise (Bialek, 2012). Nevertheless, that failure does not prove that feedback cooling cannot occur biologically, and the potential advantages encourage further investigation.

Finally, there is the need to work out in much more detail the functioning of the minicolumn. Fields et al. (2022) have made a start, but what are the specific electrical, magnetic, vibrational, and quantum mechanical processes involved?

## 11. Specific advantages

Although there are limitations to the model, we see a number of distinct advantages of the feedback cooling model, and these are now summarised and commented on.

### 11.1 Pinnacle of efferent and afferent

The elements of the cortical minicolumn model are poised just where the mind needs to be for both sensing and acting. The core units are already at that junction in their unconscious form, and they just need feedback cooling to elevate them – or evolve them – into a more highly powered quantum mechanical processor. The cortical minicolumns or psychons are in the right place to latch on to the information stream and begin to process it nonalgorithmically, alleviating a concern that Grush and Churchland (1995) had with the Penrose–Hameroff model. That means that, in evolutionary terms, the minicolumn/psychon is already in place ready for a feedback circuit to be added to it so as to achieve additional processing power, and, as an added bonus, consciousness.

The feedback cooling model is absolutely specific in pointing to the anatomical substrate of consciousness. The model suggests that feedback cooling, activated by thalamocortical circuits, switches clusters of pyramidal cells from subconscious mode into conscious mode, producing a coherent entity – a psychon in Eccles' terms or a minicolumn according to Stapp – which becomes entangled with all the other 100 million or so other units to form a Bose–Einstein condensate. Our conscious mind is essentially a powerful quantum accelerator sitting on top of a lower-level neurocomputer. Working together, the combination bestows on us extraordinary mental abilities (Penrose, 1989). Through reprogramming of the lower-level computer by the upper-level computer, we can learn novel things, whether it be acrobatics or a new language. We are able to see mathematical proofs, or perhaps reduce numbers to prime factors in our heads (as Gauss was reportedly able to do), and it gives us a glimpse into how Ramanujan and similar savants could possibly perform their feats. We might also have a handle for understanding subtler dimensions of the human mind – is meditation, for example, a case of the upper computer exploring itself? [23]

### 11.2 The feedback cooling idea is testable

As set out earlier, NMS techniques could be used to detect cold spots in the brain when it is awake. Cooler temperatures reflect the presence of bosons which take part in quantum mechanical correlations between different cortical regions (the whole cortex acting as a single superconducting sheet). Because the theory here explains the how and why of consciousness, it should be classed as a true theory, not just a description of correlations (Schurger & Graziano, 2022).

### 11.3 Sleep and wakefulness

The cooled psychon model gives a clear explanation of the difference between sleep and wakefulness, and a rational framework for describing the brain states of sleepwalkers. Crucially, sleepwalkers actually exist, unlike the fraught case of philosophical zombies, making discussion of these intriguing cases much more tractable. The model explains why consciousness provides a distinct advantage over unconscious neural processing – consciousness reflects the action of a quantum accelerator that greatly increases computing power. To give a rough idea of the difference, Stapp (1995) calculates (p. 7) that a classical system might contain M × N states and need the same number of registers to describe it, whereas a quantum description of the same system would need $2 \times (2L + 1)^{M \times N}$ registers, where L is the number of possible states in each register. This exponentially greater number lets us understand why consciousness confers a distinct evolutionary advantage, and why it has probably appeared in many creatures on multiple occasions.

---

[23] Meditational experiences seem to suggest that it may be possible to shut down the neural computer while leaving oneself immersed in the quantum computer. One thing this ancient practice demonstrates is that consciousness, like superconductivity, is either on or off – not graded in the way that Godfrey-Smith (2021, p. 262) sees it.

### 11.4 Anesthesia

As with sleepwalking, the present model provides a coherent picture of what might be happening with anesthesia. Although the mode of action of anesthetics is still not known in detail, there is good evidence that the same thalomocortical loops which give rise to 40 Hz oscillations in the cortex play a key role in regulating consciousness (Edelman & Gally, 2013). We have suggested that a major function of thalamocortical loops is to regulate the feedback cooling of minicolumns, and hence switch on consciousness, and the same process might be said to operate during anesthesia (Alkire et al., 2000; Alkire & Miller, 2005). Alkire and colleagues describe how, via the control of the thalamocortical loop, anesthesia causes firing of cells to change from tonic to burst-like, turning consciousness into unconsciousness. In later work, they elaborate on their "thalamic consciousness switch" hypothesis, and propose that anesthetics act to prevent coordinated communication between thalamus and cortex, disengaging both sensory input and motor output (Alkire & Miller, 2005).

Remarkably, the inert gas, xenon, acts as an anesthetic, and it turns out that its effectiveness seems to depend on what isotope of xenon is used (Li et al., 2018; Smith et al., 2021). That strongly implies that the atomic number of the isotope, reflecting the quantum mechanical spin of its nucleus, controls the effectiveness of the resulting anesthesia, and Li and colleagues point out that such a finding points to quantum mechanics playing a fundamental role in consciousness (Li et al., 2018, p. 271); see also Hameroff (2018). The involvement of nuclear spin provides evidence that anesthesia interferes in some way with the spins of certain nuclei in the brain, and this in turn supports the idea that consciousness involves long-lived quantum spin states, similar to how NV states in diamond also depend on spin states.

## 12 Conclusions

In 1986, before a quantum computer had even been constructed, Michael Lockwood wrote a piece entitled "Could the brain be a quantum computer?" and, based on theoretical considerations by Deutsch (see Deutsch and Jozsa (1992)), answered the question with a tentative yes (Lockwood, 1989, Ch. 14).[24] Lockwood recognised that there were things a quantum brain could do that a classical brain could not, greatly increasing its computational power.

He refers to the paper of Marshall (1989) and notes how consciousness may be associated with Bose–Einstein condensates. "Bose-condensed states," he says, "are exactly the sort of thing that is needed if the brain is to operate as a quantum computer. They lend themselves to coherent superposition, with constructive and destructive interference, in just the way that is required of quantum computer memory states" (Lockwood, 1989, p. 259). Lockwood notes that evolution is the ultimate opportunist, and "…if the possibility is there, we shouldn't baulk at the thought that natural systems may have taken advantage of it" (loc. cit.).

Lockwood later says how that the mind–body problem is not a problem to do with the mind – we are intimately acquainted with it all the time ("In awareness, we are, so to speak, getting an insider's look at our own brain activity" Lockwood (1998, p. 88)) – but with our conception of matter, which is an antiquated classical version due to Newton, not a modern quantum mechanical version with all the subtleties of waves, entanglement, and simultaneity (Lockwood, 1989, p. ix). We know all about the ghost, it's the machinery we don't understand. Lockwood (1998) repeats the words of a character in one of Sagan's novels: "Think of what consciousness feels like … Does that feel like billions of tiny atoms wiggling in place?" (ibid., p. 88).

Beginning with the seminal ideas of Eccles and Stapp, this paper has attempted to construct a direct link between long-lived macroscopic quantum mechanical states in the cortex, which arises from feedback cooling, and the insider's view of qualia. The motivation is to examine whether feedback cooling has the potential to bridge the notorious brain–mind gap. In his textbook on biophysics, Bialek says how many people tend to regard physics as stopping somewhere along the path from the outside world to subjective experience, at which point something uniquely human takes over. He relates how Helmholtz hoped otherwise, and concludes his own afterword with "so do I" (Bialek 2012, p. 471). We are really searching for a fusion of the subjective and the objective, and as Mørch (2023, p.67) notes, it's far too soon to give up.

---

[24] We have already noted the negative answer given by Tegmark (2000) based on body-temperature calculations.

Did Nature, not without precedent in such things, get there first and implement a quantum computer?

## Acknowledgements

I thank Marcus Doherty and Darryl Mathieson for helpful discussions.